\documentclass{article}
\usepackage{graphics}
\newcommand\be{\begin{equation}}
\newcommand\ee{\end{equation}}
\newcommand\eq{\begin{equation}}
\newcommand\en{\end{equation}}

\newcommand{\myskip}{\bigskip}

\newcommand{\prd}{Phys.Rev.D}
\newcommand{\prl}{Phys.Rev.Lett.}
\newcommand{\apj}{ApJ}
\newcommand{\apjl}{ApJL}
\newcommand{\apjs}{ApJS}
\newcommand{\mnras}{MNRAS}
\newcommand{\aap}{AAP}
\newcommand{\nat}{NAT}

\font\bigbf=cmbx10 at 17.28truept
\def\spose#1{\hbox to 0pt{#1\hss}}
\def\simlt{\mathrel{\spose{\lower 3pt\hbox{$\mathchar"218$}}
        \raise 2.0pt\hbox{$\mathchar"13C$}}}
\def\simgt{\mathrel{\spose{\lower 3pt\hbox{$\mathchar"218$}}
     \raise 2.0pt\hbox{$\mathchar"13E$}}}

\begin{document}

\centerline{\bigbf TACMB-1:  The Theory of Anisotropies}
\smallskip
\centerline{\bigbf in the}
\smallskip
\centerline{\bigbf Cosmic Microwave Background:}
\smallskip

\bigskip
\centerline{\bf Martin White and J.D. Cohn}
\centerline{\bf Dept. of Astronomy}
\centerline{\bf 601 Campbell Hall}
\centerline{\bf UC Berkeley}
\centerline{\bf Berkeley, CA  94720-3411  USA}
\medskip
\centerline{\bf Abstract}

This Resource Letter provides a guide to the literature on the
theory of anisotropies in the cosmic microwave background.
Journal articles, web pages, and books are cited for the following 
topics:  discovery, cosmological origin, early work, recombination,
general CMB anisotropy references, primary CMB anisotropies (numerical,
analytical work), secondary effects, Sunyaev-Zel'dovich effect(s),
lensing, reionization, polarization, gravity waves, defects,
topology, origin of fluctuations, development of fluctuations,
inflation and other ties to particle physics, parameter estimation,
recent constraints, web resources, foregrounds, observations and
observational issues, and gaussianity.

\medskip
\myskip
\noindent {\bf Introduction}
\smallskip

The cosmic microwave background (CMB) radiation is a relic of a time when
the universe was hot and dense, and as such it encodes a wealth of information
about the early universe and the formation of the large-scale structure we
see in the universe today.  The very existence of the CMB is one of the four
pillars of the hot big bang cosmology.  That the 
spectrum \cite{FIRAS96,FIRAS98,matheretal} is the best measured
black body spectrum in nature provides stringent constraints on its origin and
on any injection of energy at early times \cite{NorSmo98}.
Perhaps the most exciting and active area of CMB research, however, is the
study of its {\it anisotropies\/}: the small fluctuations in intensity from
point to point across the sky.  As we shall discuss below, these anisotropies
provide us with a snapshot of the conditions in the universe about $300,000$
years after the big bang, when the universe was a simpler place.
This snapshot is both our earliest picture of the universe and an encoding of 
the initial conditions for structure formation.

As a consequence of the hot big bang model, the CMB was predicted
for a long time \cite{Gam46,AlpBetGam48}.
The history and drama of its discovery \cite{PenWil,Dicke} 
is a full story in and of itself, starting points are the ``historical''
references \cite{MelMel94,3K}.
For the anisotropies, although several fundamental calculations were done
before 1992, it was with the COBE \cite{cobe} detection in 1992 that interest
and activity exploded.

\myskip
\noindent {\bf Origin of the CMB}
\smallskip

If we run the expansion of the universe backwards in time, the
universe becomes hotter and denser.  Beyond a point when distances
in the universe
were only $0.1\%$ of their current size, the temperature was high enough
to ionize the universe, and it was filled with a plasma of
protons, electrons, and photons (plus a few He nuclei and traces of
other species).
This transition from a neutral to an ionized medium is especially important.
Before this time the universe could be modelled as a smooth gas of photons,
baryons (the protons and electrons) and dark matter.
Since the number density of free electrons was so high, the universe was
opaque to the microwave background photons: the mean free path for photons
to Thomson scatter off electrons was extremely short.  Consequently,
the photons and baryons could be considered as a single
``tightly coupled'' fluid \cite{PeeYu70}.
In this fluid, the baryons provide the weight, while the photons provide the
pressure (for photons, pressure and density obey $p_\gamma=\rho_\gamma/3$).
As the universe expands, the wavelengths of photons are stretched out,
lowering their energy.
Eventually, when the universe had cooled to $T\sim 4,000$K, the photon
energies became too small to ionize hydrogen \cite{Pee68}
($kT \sim$ 0.25 eV, smaller than the ionization energy of hydrogen, 13.6 eV,
because the photon to baryon ratio, $10^9:1$, is so large that the high-energy
tail of the thermal distribution is significant).
At this point the protons and electrons (re)combined to form neutral
hydrogen and the photon mean free path increased to essentially the size
of the observable universe.\footnote{The mean free path of photons through
the universe must be huge or we would not see galaxies and quasars out to
distances of thousands of Mpc (1 Mpc = $3.3\times 10^6$ light years [lyr]).}
The photons were set free, and have since travelled almost unhindered through
the universe.

Once it started, the recombination of hydrogen was a phase transition,
completing very rapidly.  We refer to this time as the epoch of
recombination.
When we observe the universe in the microwave bands we see the photons
which last interacted with matter at this epoch.
These photons have travelled to us from a sphere, centered on the observer
and known as the surface of last scattering, whose radius is essentially the
entire observable universe $\sim 10^4$Mpc or $10^{10}$ light years.
The photons have continued to lose energy with the expansion of the
universe, and now form a black body with a temperature of 2.73K.
One can think of the temperature of the
cosmic microwave background photons as the temperature of
the universe.

\myskip
\noindent {\bf Describing CMB anisotropies}
\smallskip

Numerous observations of the cosmic microwave background photons support
this assumption of cosmological origin \cite{cosm-arguments}:
the background is isotropic \cite{PenWil,CorWil76,Smo77}
and a black body \cite{FIRAS96,FIRAS98,matheretal}
and has no correlations with local structures in the universe
\cite{cobe,BouJah93,Ban96,Kne97}. 
Upon closer examination the CMB temperature is not uniform across the sky,
but has slight fluctuations from place
to place.  We shall be interested in the fluctuations of the temperature
about the mean: $\Delta T(\hat{n})$ where $\hat{n}$ is a unit vector pointing
in a particular direction on the sphere.

The largest anisotropy is a fluctuation of about 1 part in 1000 that forms
a dipole pattern across the sky.  The reason for this dipole is that the
earth is not at rest with respect to the CMB, and we see a Doppler shift in
the CMB temperature owing to our relative motion.
Since this changes as the earth orbits the sun, this dipole is modulated
throughout the year.
One of the great triumphs of modern cosmology is that if we take the mass
distribution observed around us and compute from this a gravitational
acceleration, then multiply this acceleration by the age of the universe, we
obtain a good match to both the direction and the amplitude of our velocity
vector in the CMB rest frame \cite{dipole-check}.
However this dipole is clearly of (relatively) local
rather than primordial origin, and so we generally
subtract it (plus the mean or ``monopole'') before dealing with the CMB
anisotropy.  

After this dipole is taken out, the size of the fluctuations is about 1
part in 100,000.
Mathematically we describe these anisotropies by expanding the temperature
field on the sphere using a complete set of basis functions, the spherical
harmonics
\begin{equation}
{\Delta T\over T}(\hat{n}) = \sum_{\ell=2}^\infty \sum_{m=-\ell}^{\ell}
  a_{\ell m} Y_{\ell m}(\hat{n}) \; .
\end{equation}
The $a_{\ell m}$ are a curved-sky version of a Fourier transform of
the temperature field.  By definition the mean value of the $a_{\ell m}$
is zero.

As there are no preferred directions cosmologically, theories 
predict only statistical information about the
sky, not that the temperature in a certain direction should have a particular
value.  For this reason the quantities of interest are statistics of
the observed temperature pattern.
The most common and useful statistic is known as the
correlation function (or $2-$point function) of the temperature field
$C(\theta)$.  We form this by calculating the average of
$\Delta T/T(\hat{n}_1)\Delta T/T(\hat{n}_2)$
across all pairs of points in the sky $(\hat{n}_1,\hat{n}_2)$ separated by an
angle $\theta$ ({\it i.e.}~$\cos\theta=\hat{n}_1\cdot\hat{n}_2$).
Under the assumptions that our theory has no preferred direction in the sky
(statistical isotropy) and that the fluctuations in temperature have Gaussian
statistics, the correlation function encodes all of the physical information
in the CMB anisotropies.  (For non-Gaussian fluctuations there will be
additional information in higher order, {\it e.g.}~3-point, correlations.)

Original theoretical calculations and observations were performed almost
entirely on the correlation function.  However, Wilson and Silk, 
\cite{WilSil81} introduced
the ``multipole expansion,'' which isolates the physics much more robustly
and simplifies the calculations:
\begin{equation}
\label{cleq}
  C(\theta) = {1\over 4\pi} \sum_{\ell} (2\ell+1) C_\ell P_\ell(\cos\theta)
\end{equation}
where $P_\ell(\cos\theta)$ are the Legendre polynomials and the $C_\ell$ are
the quantities of interest known as the multipole moments.  In terms of the
$a_{\ell m}$ defined above
\begin{equation}
  C_\ell = \sum_{m=-\ell}^\ell  a_{\ell m}^{*} a_{\ell m}
\end{equation}
One can think of $\ell$ as the variable ``Fourier'' conjugate to angle,
$\ell \sim \theta^{-1}$.

{}From this description one of the fundamental limitations to the study of
CMB anisotropies becomes evident.  We are trying to estimate these quantities
$C_\ell$ statistically from a finite number of samples, hence our estimates
will be uncertain by an amount proportional to the square root of the number
of samples (often called ``cosmic variance'' since we would need more
universes to get a better determination) \cite{AbbWis84,ScaVit}.
Each $C_\ell$ comes from averaging over $2\ell+1$ modes, and thus the
sample variance error on $C_\ell$ is
\begin{equation}
  {\delta C_\ell\over C_\ell} = \sqrt{2\over 2\ell+1}
\end{equation}
where the $2$ in the numerator arises because $C_\ell$ is the
square of a Gaussian random variable ($a_{\ell m}$) and not the variable
itself.\footnote{The variance of $x^2$ is twice the square of that of $x$
if $x$ is a Gaussian random variable of zero mean.}.  If only a fraction
$f_{\rm sky}$ of the sky is observed then the error is increased by
$f_{\rm sky}^{-1/2}$ \cite{ScoSreWhi94,Sco95,MagHob96,JKKS96b}

\myskip
\noindent {\bf The physics of CMB anisotropies I:
The simplest picture}
\smallskip

Current CMB anisotropy measurements are improving rapidly
 \cite{experiments}
and a broad outline of $C_\ell$ as a function of $\ell$ is taking shape.
At large angular scales (small $\ell$) there is a flat plateau that rises
into a narrow peak at about one degree.  On arcminute scales the power
has fallen once more and on even smaller scales currently there
are only upper limits.  In this section we shall describe how we now
understand this structure.

Historically the key ingredients were the recognition that in the early
universe there was a tightly coupled photon-baryon plasma \cite{PeeYu70}
that decoupled suddenly \cite{Pee68,ZelKurSun69,SeaSasSco99,SeaSasSco00,
recfast}
and that in the presence of perturbations the fundamental modes of excitation
were sound waves 
\cite{Sil68,PeeYu70,DorZelSun78,WilSil81,SilWil81,BonEfs84,VitSil84}.
However the language has changed significantly since those early papers,
and the sophistication with which the calculations are performed has improved
radically \cite{HSSW,CMBFAST,CAMB},
so below we outline this more modern description, developed in
\cite{WhiScoSil94,Sel94,echoes,HuSug95a,HuSug95b}.

This description falls within the current
paradigm of cosmological structure formation \cite{LSS-review}.
When we observe the distribution of galaxies about us we find that they are
not arranged at random, but rather cluster together in coherent patterns that
can stretch for up to 100Mpc.  The distribution is characterized by large
voids and a network of filamentary structures meeting in large overdense
regions.
A great deal of evidence suggests that this large-scale structure arose
through the action of gravity on initially small 
amplitude perturbations in density.
As inflation \cite{Lid99,Tur97,LytRio99,LidLyt00} 
is the most promising theory for the origin
of these primordial density perturbations,  we will use its properties for
illustrative purposes.  Generically inflation predicts
that at very early times there were small, almost scale-invariant
adiabatic\footnote{The term adiabatic implies that a positive fluctuation in
the number density of one species is also a positive fluctuation in all of
the other species ({\it i.e.}~more photons means more baryons and more dark
matter).  A spectrum of fluctuations is scale-invariant if the gravitational
potential fluctuation it produces has the same amount of power per logarithmic
interval in wavelength.} fluctuations in the density of the universe on a wide
range of physical scales.
A region of space that was initially overdense would give rise to a larger
than usual gravitational potential.  Surrounding matter would fall into this
potential, increasing the overdensity.  Similarly matter would flow out of
regions of underdensity, increasing the density contrast further.  In this
way gravity can amplify any already existing density perturbations.
Eventually the density contrasts would become so large that we could
ignite nuclear fusion and form stars, galaxies, etc.

The CMB anisotropies that we see are a snapshot of the conditions
when the universe was 300,000 years old, that is on the surface of
last scattering, plus some (small) processing that occurred en route to us.
After last-scattering the CMB photons stream essentially freely to us and the
density fluctuations are seen as CMB temperature differences
(anisotropy) across the sky
($\delta T/T={1\over 4}\delta\rho_\gamma/\rho_\gamma$, since $\rho_\gamma
\propto T_\gamma^4$).  The key concept  is that
anisotropy on a given angular scale is related to density perturbations
on the last scattering surface of a given wavelength.  
The relevant wavelengths correspond to the length projected by that angle
on the last-scattering surface:
$\lambda \sim 200$Mpc$ \,(\theta  / {\rm deg})$.
Phrased another way, multipole moment $\ell$ receives its dominant
contribution from Fourier mode $k$, where $\ell=kr$ and $r$ is the
(comoving angular diameter) distance to last scattering.

We show in Fig.~1 a theoretical prediction for the anisotropy spectrum
\begin{figure}[h]
\begin{center}
\resizebox{3.5in}{!}{\includegraphics{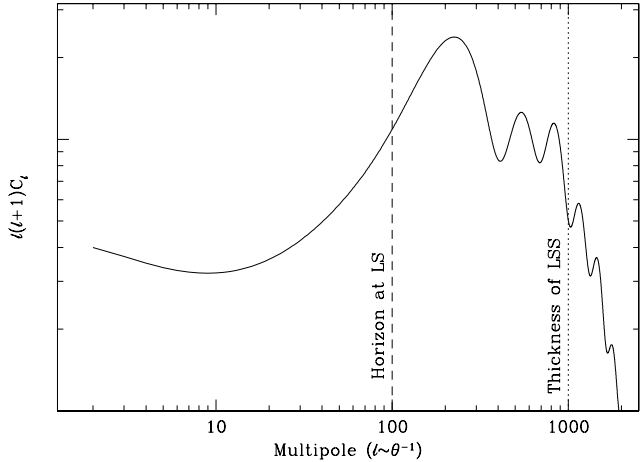}}
\end{center}
\caption{ The CMB angular power spectrum (Eq.~\protect\ref{cleq}) as a
function of multipole moment $\ell \sim 1/\theta$.  Roughly, one
degree on the sky today corresponds to $\ell \sim 10^2$, one arcminute
to $\ell \sim 10^3$.}
\end{figure}
(i.e.~$\ell(\ell +1)C_\ell$ vs.~$\ell$)\footnote{It is conventional to plot
$\ell(\ell+1)C_\ell$ rather than $C_\ell$ because this is approximately the
power per logarithmic interval in $\ell$ (or angle).  Also, in the simplest
possible model of scale-invariant fluctuations from the Sachs-Wolfe effect
(see below) $\ell(\ell+1)C_\ell$ is constant.} 
of a cosmological model with cold dark matter and an initial spectrum as
given by inflation.
Similar to the current observations mentioned earlier, the spectrum clearly
has 3 distinct pieces: at low-$\ell$ (large angular scales) there is a flat
plateau that rises into a series of bumps and wiggles that then damp
quasi-exponentially on small angular scales.
These 3 regimes are separated by 2 angular scales, the first at about
1 degree and the second at a few arcminutes.

To understand the origin of these features let us go back in time to just
before recombination.  At this time the universe contained the tightly
coupled photon-baryon fluid and dark matter, with perturbations in the
densities and thus gravitational potentials on a wide range of scales.
While perturbations in the dark matter grow continuously as the universe
ages, the gravity-driven collapse of a perturbation in the baryon-photon
fluid is resisted by the pressure restoring force of the photons.
For example, as an overdensity falls into a gravitational potential it
becomes more and more compressed.  Eventually photon pressure halts the
collapse and the mode rebounds, becoming increasingly rarefied.  The expansion
is slowed and halted owing to the weight of the fluid and the gravitational
potential, causing the mode to recollapse once more.
In short, an acoustic wave is set up, with gravity the driving force and
pressure the restoring force.  Mathematically, the Fourier mode $k$ of the
temperature fluctuation is governed by a harmonic-oscillator-like equation
\cite{PeeYu70,HuSug95a,HuWhi96a}
\begin{equation}
\left[ m_{\rm eff} \Delta T_k' \right]' + {k^2 \over 3}\Delta T_k
  = -F_k
\label{eqn:Oscillator}
\end{equation}
where $F$ is the gravitational forcing term owing 
to the dark-matter potentials,
$m_{\rm eff}$ describes the inertia of the fluid,
and primes denote derivatives with respect to (conformal) time
($\eta = \int dt/a(t)$ where $a(t)$ is the scale-factor of the universe).
The solutions are acoustic waves.

We are now in a position to understand the features in Fig.~1.

The large-angular scale (Sachs-Wolfe) plateau ($\ell < 100$) in the angular
power spectrum arises from perturbations with periods longer than the age of
the universe at last scattering, {\it i.e.} $\sim$ larger than the horizon,
scales that can be affected by causal physics at that time.  
These waves are essentially frozen in their initial configuration and provide
us with a probe of the physics that created them, unspoiled by cosmological
evolution.
Since CMB photons lose energy climbing out of the potential wells associated
with these long-wavelength density perturbations, the temperature differences
seen on the sky reflect the gravitational potential differences on the
last-scattering surface \cite{SacWol67,Pea91,WhiHu97}.
If the density fluctuations are approximately scale-invariant the plateau in
the angular power spectrum is flat.

At scales smaller than the horizon, the baryon--photon fluctuations that
produce anisotropy on sub-degree angular scales ($10^2<\ell<10^3$) have
sufficient time to undergo oscillation.
At maximum compression (rarefaction) the CMB temperature is higher (lower)
than average.  Neutral compression corresponds to velocity maxima of the
fluid, which leads to a Doppler-shifted CMB temperature.  The Doppler effect
is  subdominant because we see only the line-of-sight component of the
velocity and the speed of sound is less than the speed of light.
Since last-scattering is nearly instantaneous, the CMB provides a snapshot of
these acoustic oscillations, with different wavelength modes being caught
in different phases of oscillation.  Because a given multipole $\ell$ is
dominated by the effects of a narrow band of Fourier modes, this leads to
peaks and valleys in the angular power spectrum.
The peaks are modes that were maximally under or overdense at
last-scattering (since the power spectrum is the amplitude {\it squared}), 
and the troughs are velocity maxima, which are $\pi/2$ out
of phase with the density maxima. 

On even shorter scales ($\ell \simgt 10^3$) the finite duration of
recombination has an observable effect \cite{Sil68,EfsBon87}.
During this time the photons can random walk a distance given by
the mean free path (which is increasing during recombination) times the
square root of the number of scatterings
\cite{HuSug95a,HuSug95b,HuWhi97-damp}.
Thus photons can diffuse out of any overdensity
on smaller scales than this.  This leads to an exponential damping of the
spectrum on small scales (known as Silk damping).
If we approximate last scattering as extremely rapid, the damping is
exponential with e-folding scale the geometric mean of the horizon and the
photon mean free path.
The finite duration of last scattering changes this somewhat, and the
damping is closer to an exponential of a power of scale
\cite{HuSug95a,HuSug95b}.

\myskip
\noindent {\bf The physics of CMB anisotropies II:
Beyond the simplest picture}
\smallskip

While the above picture explains the gross features of Fig.~1, a number
of other effects have received detailed study.  Here we discuss these
effects in the order in which they occur in the evolution of the universe,
which is not the historical order in which they were discovered.

At last scattering and since, the photons not only respond to the gravitational
potentials caused by dark matter density perturbations, but also to
any other perturbations in the space-time metric.  Technically,
gravitational potentials owing to density perturbations are often referred to as
scalar, corresponding to their Lorentz transformation properties
(properties under boosts and rotations).
Since the metric has more complicated transformation properties
(specifically it is a spin-2 tensor), 
vector and tensor fluctuations are also possible.
Vector perturbations, also called vortex perturbations, decay as the
universe expands unless they are constantly generated.  
Tensor perturbations (also called gravity waves) can be generated by quantum
fluctuations of the spacetime.  These persist and can have an effect in
many cases.  Tensor perturbations of spacetime do not create the same
baryon-photon oscillations, but can contribute a Sachs-Wolfe plateau
\cite{SacWol67,Sta79,Sta83,FabPol83,Sta85,AbbWis84,AbbSch86,Whi92,CBDES,
HuWhi97}.
Inflationary theories usually produce no vector perturbations, and small
tensor perturbations.

As the photons travel through the universe from the surface of last scattering
they can interact gravitationally with the matter.  If the gravitational
potentials are still evolving, additional temperature perturbations are
generated by the ``integrated Sachs-Wolfe  effect''
\cite{SacWol67,KofSta85,AbbSch86}.
Schematically a photon falling into a gravitational potential will gain
energy.  If the potential evolves during the photons' traverse, the energy
lost climbing back out will be different from that gained falling in, leading
to a net anisotropy.
To linear order in the perturbations the gravitational potential
$\phi$ is constant when matter dominates the energy budget of the universe
and this phase gives no contribution.
However, right after recombination photons still contribute enough to the
energy density of the universe that the change in time
of the potential, $\dot{\phi}$, is non-zero (the ``early ISW effect'') and
at very late times if either curvature or a cosmological constant dominate
$\dot{\phi}\ne 0$ (the ``late ISW effect'').
Additionally when non-linear structures form the potential can change with
time owing to both the growth and movement of bound halos leading to
anisotropies through the Rees-Sciama effect \cite{ReeSci68}.
In modern theories this effect is very small, and is not the dominant source
of anisotropy on any scale \cite{Sel96a,TulLagAnn96,PynCar96}.

In addition to the energy gained and lost by photons, the path a photon takes
is altered by non-zero potentials.  This gravitational lensing causes the
spectrum to be slightly ``blurred''
\cite{ColEfs89,Lin90a,Lin90b,CayMarSan93a,CayMarSan93b,Sel96b,MetSil97},
smoothing the third acoustic peak by a few percent and slightly altering
the shape of the damping tail.
The signature of gravitational lensing may be used to reconstruct the
projected gravitational potential along the line-of-sight
\cite{ZalSel99}.

Observations of the spectra of high redshift QSOs indicate that the universe
is highly ionized out to redshift
$z\sim 5$.  Thus photons can again scatter off free
electrons in a second ``scattering surface.''  Unlike the $z\sim 10^3$ surface,
however, the electron density today is quite low, and the baryons and photons
do not become tightly coupled.  Because of this the two fluids can have a
large relative velocity, which enhances the power of the Doppler effect.
Reionization, as this is called, damps power on angular scales smaller than
the horizon subtended by the epoch of reionization
while generating extra power owing to Doppler scattering
\cite{Kai84,EfsBon87,SugSilVit93,HuScoSil94,DodJub95,HuWhi96b,HuWhi97-damp}.
There is also a second order effect known as the Ostriker-Vishniac effect
\cite{OstVis86,Vis87,HuWhi96b,JafKam98,Hu00} that affects only the small
scale.

It is unlikely that the reionization of the universe will occur uniformly
throughout space, so anisotropies will be generated owing to the ``patchiness''
of reionization.  Depending on the redshift of reionization and whether the
ionizing sources are quasars or stars, the angular scale of this anisotropy
could be quite different.  Early analytic attempts to discuss patchy
reionization \cite{ADPG96,GruHu98,KnoScoDod98,Hu00} 
used crude models to
estimate the required correlation functions.  Recent numerical
simulations \cite{NorPasAbe,BruFerFabCia00,BenNusSugLac00,GneJaf00}
have improved upon these results, but this remains an area of active
research at present.  Current calculations suggest the patchy reionization
will not dominate except on extremely small angular scales.

Finally, once structure formation is well underway, the photons can interact
with hot gas in the intergalactic medium \cite{SunZel72,SunZel80}.  The CMB
photons can either be upscattered in energy when interacting with hot gas
(the "thermal" Sunyaev-Zel'dovich effect) or have their temperature altered
by Doppler scattering from moving gas (the "kinetic" S-Z effect).
For recent reviews see \cite{Rep95,birkinshaw}.
The thermal SZ effect is probably the largest source of anisotropy on
angular scales of a few arcminutes and has been calculated both analytically
\cite{KomKit99,HolCar99,Coo00,MolBir00,ZhaPen00}
and numerically
\cite{daSBarLidTho00,RKSP,SelBurPen00,SprWhiHer00,daSBarLidTho}

On top of these effects are ``foregrounds'' (as the signal is a background)
that mask the CMB physics and are the source of many headaches and much work.
These include dust, free-free emission, and synchrotron radiation, all of which
have estimated dependencies on frequency and angular scale (a summary and
comparison of these can be found in \cite{MAFbook,maxfor}, some
references are \cite{TegEfs96,fgndsias,Kno99,Ref00}).
Many experiments measure the CMB in many different frequency bands 
to account for these foregrounds.  Some foregrounds, such as
point sources, will produce non-gaussian anisotropies.
As the simplest inflationary models produce gaussian anisotropies, this
also can be used to distinguish them from the desired signal.  

\myskip
\noindent {\bf Polarization}
\smallskip

Not only do the CMB photons have temperatures, as described above,
they also are expected to have polarization \cite{Ree68,Kai83,BonEfs87}.
The Thomson scattering cross section $\sigma$ as
a function of solid angle $\Omega$ depends on polarization
\begin{equation}
{d\sigma\over d\Omega} \propto \left| \varepsilon_i\cdot\varepsilon_f\right|^2
\end{equation}
where $\varepsilon_{i,f}$ are the incident and final polarization directions.
The scattered radiation intensity peaks normal to, and with polarization
parallel to, the incident polarization.  If the incoming radiation field is
isotropic then orthogonal polarization 
states balance and the outgoing radiation
remains unpolarized.  In the presence of a quadrupole anisotropy, however, a
linear polarization is generated by scattering.

Since we have observational evidence for anisotropies at last scattering,
we expect that the CMB be linearly polarized.  The degree of polarization is
directly related to the quadrupole anisotropy at last scattering.
While the exact properties of the polarization depend on the mechanism for
producing the anisotropy, several general properties arise.
The polarization peaks at angular scales smaller than the horizon at
last scattering ({\it i.e.}~smaller scales than the first temperature peak) 
owing to causality.
Since only those photons that scattered in an optically thin region near
last scattering could have had a quadrupole anisotropy, the polarization
fraction is small and dependent on the duration of last scattering.
For the standard thermal history it is a few percent of the temperature
anisotropy.  An additional change in polarization can occur during
subsequent interaction with ionized matter (e.g.~during reionization 
as mentioned above \cite{HogKaiRee82}).
Gravitational interactions do not generate or destroy polarization.

The formalism for the description of polarized radiation on the sphere has
been developed in \cite{SelZal,KamKosSte,ZalSel,TAMM}.  In analogy with the
temperature, the polarization is expanded in a series of spin-weighted
spherical harmonics whose coefficients can be used to define ``E-mode'' and
``B-mode'' polarization power spectra\footnote{The modes are called ``E''
and ``B'' to denote their parity transformation properties; they should not
be confused with the electric and magnetic fields of the CMB signal itself.
Some authors also refer to these as the ``gradient'' and ``curl'' components
in analogy with the decomposition of a vector field.}
that transform into one another under a 45-degree rotation of the
polarization.  There is additionally a cross-power spectrum between T and E.
Density (or scalar) perturbations have no ``handedness'' and so generate only
E mode polarization.
Vector and tensor modes create both E and B mode polarization.

Information from polarization is complementary to information from
temperature anisotropies.
Different sources of anisotropy (scalar, vector, tensor) generate
different patterns of polarization \cite{CriDavSte93,CriCouTur95,TAMM}
and adiabatic and isocurvature modes generate different polarization spectra
\cite{HuSpeWhi97,SpeZal97,TAMM}.
The presence of polarization increases the number of spectra that can be
measured from 1 to 4 (temperature, the two polarizations, and the T-E cross
spectrum), which allows better constraints on cosmological models
\cite{Sel97,ZalSpeSel97}.
More beneficially, polarization depends on some of the cosmological
parameters differently than the temperature anisotropy, allowing degeneracies
in the fitted parameters to be removed and improving parameter constraints
by a large factor \cite{Sel97,ZalSpeSel97,EisHuTeg98,EfsBon99}.

\myskip
\noindent {\bf What can we learn from CMB anisotropies?}
\smallskip

CMB anisotropies represent one of the cleanest astrophysical systems known:
the anisotropies arise from electron-photon interactions and weak
gravitational fields.  Thus the predictions can be calculated accurately
and reliably, while at the same time providing us with valuable information
about the early universe, the formation of large-scale structure, and the
cosmological parameters.  Here we discuss what we have already learned and
what we hope to learn soon \cite{Romans} from a comparison of these
calculations with high-precision observations.

To begin with generalities, because the large angle anisotropies are 1
part in $10^5$, this constrains the amplitude of the fluctuations in matter
densities on the scale of the horizon.
Any theory of fluctuation generation and evolution must be normalized to agree
with this value \cite{WhiSco96a,BunLidWhi96,GRSSB,BunWhi97,RSBG}.
Within the limited statistics currently available the fluctuations appear to
be Gaussian \cite{Kog1,BanZarGor99,BroTeg99,PhiKog}, 
as predicted by the simplest models of inflation.

Comparing the size of these early fluctuations to the size of density
perturbations today provides more circumstantial evidence for nonbaryonic
dark matter.
(Nonbaryonic dark matter was first introduced in other contexts for
other reasons.)
While a model-independent statement is difficult to make, if there were only
baryons, the level of inhomogeneity required to produce the observed
large-scale structure through gravitational infall would generally lead to CMB
anisotropy that is about ten times larger than that observed 
\cite{WilSil81}.\footnote{To get enough large-scale structure power but
suppress the large-angle anisotropies there are models with a power-law
initial spectrum whose power grows with decreasing length scale.
However, then reionization is required to flatten the spectrum at
COBE scales.  This simultaneously damps the power at degree scales, leading
to conflict with observations \cite{HuBunSug95}.}
A model based on gravitational amplification of initially small adiabatic
perturbations in a universe whose dominant matter component is cold and dark
manages to reproduce the amplitude of the fluctuations required over many
decades in linear scale.  

A first acoustic peak has been detected in the CMB temperature anisotropies
\cite{ScoWhi94,DodKno,KnoPag,Jafetal00}.
The position of this peak is related to the size of the horizon at last
scattering and the distance travelled by the photons since this time.
This angular scale is sensitive to spatial curvature
\cite{DorZelSun78,Wil83,GouSugSas91,KamSpeSug94,HuWhi96},
appearing smaller if the universe is open (negatively curved) and larger if
the universe is closed (positively curved, similar to a sphere).
The currently observed position is consistent with the universe being spatially
flat.

The amplitude of the peak indicates that either the baryon density is high,
or the the matter density of the universe is below critical (or both)
\cite{BoomerangReturns,BoomInterp,Jafetal00,TegZal00}.
Because we see any CMB signal at all on degree scales means that the
photons were able to travel unhindered to us for some time before reionization
occurred, 
{\it i.e.}~the universe was neutral for a while between $z\sim10^3$ and
$z\sim 5$ \cite{ssw,GriBarLid99,BoomInterp}.

Many of these above features cause difficulties for non-inflationary theories
of structure formation.  Of the dozens of theories proposed before 1990, only
inflation and cosmological defects survived after the COBE announcement, and
only inflation is currently regarded as viable by the majority of cosmologists.
Cosmological defects are configurations in spacetime of some field, 
{\it e.g.}~domain walls, strings, monopoles, 
or textures, which can be produced as
the universe cools through several phase transitions \cite{VilenkinBook}.
In contrast to inflation, perturbations caused by defects form continuously,
as larger and larger regions come into causal contact and feel the influence
of the defects moving around.
Calculations with defects are extremely challenging technically, making it
difficult to draw robust conclusions.  However, several trends emerged early
on: defect theories fail to reproduce the observed power in the matter
fluctuations when normalized to the CMB \cite{WhiSco96a,AlbBatRob97},
generically produce non-Gaussian fluctuations on degree scales
\cite{CouFerGraTur94,Tur96-ng}
that are not observed \cite{Kog1,BanZarGor99,BroTeg99,PhiKog},
a high redshift of reionization \cite{PenSpeTur94}, and even in
the absence of reionization give a very low (or absent) broad peak
\cite{Alb96,ACFM96} around degree scales.
In the many cases where detailed temperature anisotropy calculations have been
carried out \cite{PenSelTur97,TurPenSel98}, defect models strongly disagree
with observations.

Because the surface of last scattering is a sphere of radius $\sim 10$Gpc,
it is sensitive to any non-trivial topology in the universe.
Current measurements indicate that the large-scale structure of space-time
appears topologically simple
\cite{SteScoSil93,deOSmoSta96,CWRU-guys,ScaLevSil99}.

More information is hinted at in the current data but not yet as
precisely measured.
For example, the narrowness of the first peak means perturbations were created
a long time ago, for instance laid down early on by inflation
\cite{BoomerangReturns}.
The spectrum of initial perturbations is close to scale-invariant.
The position of the damping tail provides a feature in the power spectrum
that is almost independent of the source of the fluctuations, depending only
on the properties of the fluid at last scattering ({\it e.g.}~the 
baryon-to-photon
ratio) and the angular diameter distance to last scattering
\cite{HuWhi96b,HuWhi97-damp}.

Future determinations of the temperature spectrum and detections of the
polarization spectrum can provide even more information.
The low-$\ell$ shape and relative amplitudes of the polarization power
spectra indicate which modes (scalar, vector or tensor) are populated
by the source of fluctuations \cite{TAMM}.
The relative positions and heights of the peaks can provide a test of
inflation \cite{HuWhi96a,Tur96,HuSpeWhi97} or more generally of an apparently
acausal generation of curvature perturbations on super-horizon scales
\cite{Lid95}.

If all of our modelling assumptions are borne out, and the angular power
spectrum is well fit by an inflationary CDM model, then we can expect to
constrain on the order of $10$ cosmological parameters to the few percent level
{}from high resolution anisotropy observations, {\it e.g.},
\cite{JKKS96a,JKKS96b,ZalSpeSel97,BonEfsTeg97,Whi98,maxsn,EisHuTeg98,
EisHuTeg99,EisHuTegWhi99}.
Writing the initial spectrum as a power law, more precise
constraints on both the
power and deviations from power law behavior will be possible 
\cite{Lyt96,CovLyt99}. 

Generally the CMB constrains quite well the angular diameter distance to
last scattering, the physical matter ($\Omega_{\rm mat}h^2$) and baryon
densities ($\Omega_B h^2$), and the spectral index of the fluctuations.
Here we have written the Hubble constant as
$H_0 = 100h{\rm km}\;{\rm s}^{-1}\;{\rm Mpc}^{-1}$
and the densities as a fraction of the critical density
($\rho_{\rm crit}\equiv 3H_0^2/(8\pi G)$),
$\Omega_i=\rho_i/\rho_{\rm crit}$.
Typically some combinations of the other parameters are well constrained
while some are very poorly constrained \cite{EfsBon99}.
To break these parameter degeneracies one needs to include measurements
that complement the CMB constraints.  For example, the degeneracy between
$\Omega_{\rm mat}$ and $\Omega_\Lambda$ that enters into the angular
diameter distance to last scattering can be broken with low redshift
measurements \cite{Whi98,EisHuTeg98}.
An accurate measurement of the Hubble constant $H_0$, combined with the
accurate determination of $\Omega_{\rm mat}h^2$ mentioned above, also breaks
many degeneracies, and this is what large-scale structure surveys or direct
$H_0$ measurements can provide.

If we are lucky enough that inflation takes place near the GUT scale, then
a measurably large gravitational wave component to the anisotropy is
predicted
\cite{Sta79,Sta83,FabPol83,Sta85,AbbWis84,AbbSch86,Whi92,CBDES,HuWhi97,lyth97}.
Using both temperature and polarization information, tensor signals as
small as $0.1-1$\% \cite{Kin98,KamJaf00,JafKamWan}
of the total anisotropy can be detected, corresponding to
$E_{\rm inf}>10^{15}$GeV.
Should our luck hold out, and inflation be dominated by a single scalar
field, it may even be possible to reconstruct the inflationary potential
to some extent from detailed measurements of the scale-dependence of the
signals \cite{Lid97}.

\myskip
\noindent {\bf Nuts and bolts: Calculating CMB anisotropies}
\smallskip

While the CMB anisotropy description above is physically clear, it is
very heuristic in comparison to how the calculations are done in practice.
One begins with the coupled Einstein, fluid, and radiative transfer equations,
expanding about an exact solution and truncating the expansion at linear order
\cite{PeeYu70,Wil83,Bar80,KodSas84,BonEfs87,MukFelBra92,MaBer95,WhiSco96b,TAMM}.
This is consistent as the observed fluctuations are small; in addition, 
the higher order terms have been calculated and shown to be small as expected
\cite{HuScoSil94,DodJub95}.
This results in a set of coupled ODEs that describe the evolution of
each independent Fourier mode (or its curved-space generalization).
While in some cases an analytic solution is possible, the equations are
usually numerically integrated from early times until the present
\cite{WilSil81,Wil83,VitSil84,BonEfs87,Sto94,Sug95,MaBer95,WhiSco96b,
SelZal96,CMBFAST}.

The formalism for computing the $C_\ell$ (or the higher-order moments)
for any FRW space-time and any model of structure formation exists
 \cite{HSWZ,cam1} and for many cases of interest can be done with
publicly available codes such as CMBFAST \cite{CMBFAST,SelZal96} and 
CAMBfast \cite{CAMB,CambridgeGuys}.
These codes incorporate many refinements \cite{HSSW} and have become quite
complex.  However, in addition to calculating self-consistency within a given
code,  calculations have been done (mostly for CDM models) using several
independently developed codes, with an agreement found of ${\cal O}(1\%)$.
\cite{HSWZ}.

\myskip
\noindent {\bf Observational Outlook}
\smallskip

Since COBE first detected anisotropies, there has been a flurry of
observational ``firsts'' in CMB research.  We now have observational evidence
for a nearly scale-invariant low-$\ell$ plateau, a peak in power on degree
scales and a subsequent fall in power (``damping'') on arcminute scales.
As of this writing (Fall 2000), polarization has not yet been detected.
Improved ground-based, balloon, and satellite CMB experiments are underway
or under construction that will measure a range of properties, from
small scale anisotropies on small regions of sky to full sky maps, with
and without polarization.
The most current information in this rapidly progressing area
can be found on the experimental web pages \cite{experiments}.
Recent experiments span a larger range of angular scales than ever before,
allowing features in the spectrum to be identified from individual
experiments rather than statistical compilations.
This minimizes the effect of calibration uncertainties that can offset
different experimental results by of order 10-20\% in amplitude.
Increased sky coverage (to allow calibration off the dipole) and better
control of systematics are reducing this uncertainty in the next generation
of experiments.
Although many early measurements were statistical detections, in current
experiments the signal-to-noise on each resolution element is larger than one.
The flood of new, higher-resolution, higher signal-to-noise data has required
the development of specialized analysis tools to extract the maximum
cosmological information.  CMB analysis is a flourishing sub-field that we
have not attempted to address here.

\myskip
\noindent {\bf Summary}
\smallskip

Anisotropies in the CMB are one of the premier probes of cosmology and
the early universe.
Theoretically the CMB involves well-understood physics, in the ``linear
regime'' and is thus under good calculational control.
Model independent constraints on the cosmology and the model of
structure formation exist.
Within any given model parameter extraction can be made very precisely,
especially when CMB data is combined with other data (``complementarity'').

\myskip
\noindent {\bf Acknowledgements}

\smallskip
We thank Douglas Scott for helpful comments.
This research has made use of NASA's Astrophysics
Data System Abstract Service.
M.W.~was supported by a Sloan Foundation Fellowship and by a grant from the
US National Science Foundation.  J.D.C.~was supported by
NSF-AST-0074728.

\end{document}